**Colossal dielectric constant in high entropy oxides**

By *David Bérardan\**, *Sylvain Franger*, *Diana Dragoe, Arun Kumar Meena and Nita Dragoe\**

[*]     Dr. David Bérardan, Prof. Nita Dragoe
E-mail: david.berardan@u-psud.fr, nita.dragoe@u-psud.fr
ICMMO (UMR 8182 CNRS), Université Paris-Sud, Université Paris-Saclay, 91405 Orsay, 91405 (France)

Prof. Sylvain Franger, Dr. Diana Dragoe, Mr. Arun Kumar Meeta
ICMMO (UMR 8182 CNRS), Université Paris-Sud, Université Paris-Saclay, 91405 Orsay, 91405 (France)

Entropic contributions to the stability of solids is very well understood and the mixing entropy has been used for forming various solids, for instance such as inverse spinels.[1] A particular development was related to high entropy alloys[2, 3] (for recent reviews see ref 4 and 5) in which the configurational disorder is responsible for forming simple solid solutions and which are thoroughly studied for various applications especially due to their mechanical properties[6, 7] but also electrical properties[8], hydrogen storage[9], magnetic properties.[10] Many unexplored compositions and properties still remain for this class of materials due to their large phase space.

In a recent report it has been shown that the configurational disorder can be used for stabilizing simple solid solutions of oxides, which should normally not form solid solutions,[11] these new materials were called "entropy-stabilized oxides". In this pioneering report, it was shown that mixing five equimolar binary oxides yielded, after heating at high temperature and quenching, an unexpected rock salt structure compound with statistical distribution of the cations in a face centered cubic lattice. Following this seminal study, we show here that these high entropy oxides (named HEOx hereafter) can be substituted by aliovalent elements with a charge compensation mechanism. This possibility largely increases the potential development of new materials by widening their (already complex) phase space. As a first example, we report here that at least one

HEOx composition exhibits colossal dielectric constants, which could make it very promising for applications as large-k dielectric materials.

1. **Introduction**

Configurational entropy as driving force for stabilizing, at a given temperature, different structures has been used in several studies to tailor new materials. When this entropy is large enough, *e.g.* when mixing four or five metals like in high entropy alloys[4, 5] or five cations in binary oxides,[11] a high symmetry structure is formed at high temperature (> 850°C for the case of a mixture of five binary oxides). At high temperature this mixture forms a solid solution with rock salt structure which can be quenched and is (meta)stable at room temperature due to the small diffusion coefficients. This solid solution of binary oxides was somehow unexpected since several of the oxides used (Mg, Ni, Co, Cu and Zn oxides) *do not exhibit solid solutions* in their binary phase diagrams. For the case discussed previously the solid solution can be written as (Mg,Ni,Co,Cu,Zn)O in a face centered cubic (FCC) structure. Following this pioneering report,[11] one of the first questions to be answered is whether "classical" substitutions can be performed on these materials, thus increasing significantly the possibilities of synthesis of new materials and therefore of their properties and potential applications. Additionally the study of the physical properties of these new materials is of great interest, as they cannot be easily "guessed" from the crystal structure and the nature of the constituting elements. For example, for the parent compound (Mg,Ni,Co,Cu,Zn)O (noted hereafter HEOx-0) one may intuitively estimate that it is an insulator as all binary oxides composing it. However the band structure is probably not that of a statistical average of the five binary oxides, and it is not clear whether electronic or magnetic correlations can be maintained despite the chemical disorder.

## 2. Experimental

*Synthesis*: All samples were synthesized from binary oxides and carbonates, MgO, $Co_3O_4$, $Ni_2O_3$, CuO, ZnO, $Li_2CO_3$, $Ga_2O_3$. 2.7g of the starting powders in stoichiometric amount were mixed by mechanical grinding using a Fritsch "Pulverisette 7 Premium line", with agate balls and vials at 250 rpm during 60 min. The resulting mixtures were then pressed into 12x3x3 $mm^3$ pellets under 250 MPa, which were then heated for 12h at 1000°C under air followed by air quenching. The geometrical density of the samples was in the 75-80% range.

*XRD*: Room temperature X-ray diffraction characterizations were performed using a Panalytical X'Pert diffractometer by using a Cu-Ka1 radiation, with a Ge(111) incident monochromator and a X'celerator detector. Structure refinements were performed using FULLPROF software.[24]

*UV-VIS*: Room temperature optical diffuse reflectance measurements were performed on finely ground powders using a Varian Cary 5000 double-beam, double monochromator spectrophotometer, with a PTFE integrating sphere. The obtained diffused reflectance spectra were converted using Kubelka-Munk relation and normalized to 100%-absorbance at high energy for qualitative comparison.

*XPS*: XPS was performed on a Thermo Scientific K-alpha instrument equipped with a monochromatic focalized Al-Ka source (1486.7 eV) using a spot of 400 μm and a hemispherical analyser, at a take-off angle of 0°. Wide scan spectra were acquired at a pass energy of 200 eV and 1 eV energy step while the narrow scans were acquired with a pass energy of 20 eV or 50 eV and 0.1 eV energy step. The acquisition and interpretation of the spectra were performed using Thermo Avantage software, considering a Shirley background and peak shapes were 30%/70% L/G. The full width at half maximum for the Li line was 0.7 eV. The pressure in the analysis chamber was in the low $10^{-9}$ mbar range. Charge compensation was done by means of a "dual beam" flood gun. The samples were cleaned with Ar+ ions at 200 eV and measured with and without electron flood gun. The homogeneity of samples was checked by measurement on up to 4 different points. The calibration was done relative to C1s peak located at 284.8 eV.

*Transport properties*: Transport properties were measured with a LCR bridge Hameg 8118 (for low frequencies) with a laboratory made system in a Janis cryostat. Scan impedance analysis was performed with a 7260 impedance meter (MMates), by scanning samples from 2.3 MHz to 1.3 Hz with 20 measurements per decade and applying an AC signal amplitude of 500 mV peak to peak. The data were analysed with ZView/Zplot package (Scribner), using the Phase Inflexion Detection (PID) technique for decorrelating relaxation phenomena.

Samples used for measurements were as obtained bars, thinned to about 1 mm, and with a surface of about 20 to 30 mm$^2$. The large parallel faces were painted with silver paste for making electrical contacts. The contact resistance was of the order of a few Ohm, several orders of magnitude smaller than the resistance exhibited by the samples. For high frequency impedance measurements, sample was inserted into a Swagelok type symmetrical cell and placed into a climatic chamber (Binder) at a controlled/desired temperature.

## 3. Results and Discussion

Although the binary precursors used in our study were not the same as the ones used in the previous pioneering paper[11] where only $M^{2+}O$ powders were used, the HEOx-0 sample obtained after air-quenching from 1000°C was single phase with a rock-salt crystal structure (see Supporting Figure 1). Therefore, it seems that the initial valence state of the cations is not of primary importance for the materials synthesis in this case: a formation equilibrium can occur at high temperature with oxygen losses, and the phase obtained at high temperature will be preserved by quenching. The lattice parameter obtained from the refinement of the XRD pattern is 4.2277(2) A, which is similar to the value reported by Rost *et al.* It suggests that a reduction of the 3+ cations used as starting materials occurred during the thermal treatment (a 2+ valence state for all cations was confirmed by XPS, see later).

At a first glance, as the addition of a sixth element to the compound should obviously increase the entropy contribution, partial substitution by aliovalent elements in order to tune the electrical properties of the materials should be possible, as in the case of high-entropy alloys where the valence electron count can be tuned by doping.[12] Therefore, various substitutions by several aliovalent elements were attempted (all compositions and corresponding XRD patterns are summarized in Supporting Figure 1 and Table 1). It can be seen that the situation is more complex than expected. All samples containing a few percent of 3+ or 4+ doping elements exhibit large amounts of secondary phases. This is independent of the valence state of the doping element (3+ or 4+), of its size ($r(In^{3+})>r(M^{2+})$ whereas $r(Ga^{3+})<r(M^{2+})$)[13] or of its electronic configuration ($d^{10}$ with $In^{+3}$ or $d^0$ with $Ti^{4+}$). We note that these 3+ or 4+ cations exist regularly in octahedral configuration in binary oxides and thus would not have to be "forced" to adopt this geometry in the targeted rocksalt structure. It shows that the entropy of configuration with five cations is not sufficient to stabilize a high-symmetry solid solution in this case, at least not at 1000°C. Oppositely, the partial substitution of cations in HEOx-0 by $Li^+$ leads to single phase rock-salt samples with a decrease of the lattice parameter when increasing the Li fraction that follows a Vegard law (**Figure 1**), up-to a $Li^+$ fraction as large as 16.6%. This composition corresponds to (Mg,Co,Ni,Cu,Zn,Li)O, which evidences that $Li^+$ can be successfully introduced into the rock-salt structure. This decrease of the lattice parameter indicates that $Li^+$ does not occupy insertion sites but rather substitutes divalent cations. However, this seems inconsistent with the ionic radius of $Li^+$ being slightly larger (in octahedral environment) than the average atomic radius of the 5 divalent cations.[13] Therefore, it is indicative of other evolutions such as a charge compensation mechanism that would occur to compensate the difference of valence states between $Li^+$ and $M^{2+}$. This charge compensation could occur either through a change in the valence state of one or several of the $M^{2+}$ cations, with the formation of (most probably) $Co^{3+}$ or $Ni^{3+}$, or through the formation of oxygen vacancies, both leading to a decrease of the lattice parameters. To confirm this observation, a co-substitution by a 1+ *and* a 3+ element has been attempted, $(MgCoNiCuZn)_{1-2x}Li_xGa_xO$. The resulting sample crystallizes in a single phase FCC structure with slightly smaller lattice parameter than the pristine sample (Figure 1).

Therefore a self-compensation mechanism between the two dopants enables the partial substitution of 2+ elements by pairs of 1+ and 3+. As mentioned above, the decrease of the lattice parameter observed with Li$^+$ substitution follows a Vegard law up to x(Li$^+$)=16.6%. Above this value, the lattice parameter keeps decreasing but with a smaller rate. This suggests that at least two compensation mechanisms are present in the (MgCoCuNiZn)$_{1-x}$Li$_x$O series, depending on the Li fraction in the material. Besides, it shows that the structure is very robust against substitution by +1 elements, which can be introduced in large amounts. In the previous report,[11] it was shown that the presence of 5 divalent cations was required to stabilize the HEOx phase. A straightforward development to check the validity of the charge balance condition was to completely substitute one of the divalent cations by a combination of a +1 and a +3 cations. The resulting compound, (MgCoNiCu)$_{0.8}$(LiGa)$_{0.2}$O, also crystallizes in a FCC structure, Figure 1, with only faint amount of secondary phases and a significant decrease of the lattice parameters (the average ionic radius of Li$^+$ and Ga$^{3+}$ is smaller than the average ionic radius of (Mg, Co, Ni, Cu, Zn)$^{2+}$ ), which confirms that a divalent cation can be completely substituted by a charge balanced combination of +1 and +3 (and probably larger than +3) cations.

Collectively, these results indicate that the stabilization of (M$_5$)O high entropy oxides and the partial substitution of the cations require a good charge balance between the cations and oxygen. Moreover, they show that intrinsic charge compensation mechanisms can occur that stabilize the partial substitution of 2+ cations by 1+ cations, whereas this compensation does not occur for the stabilization of the charge in the case of substitution by 3+ cations. Last, they show that besides the large variety of combinations of 2+ cations that could lead to the stabilization of (M5)O, with probably numerous different physical properties and potential applications, the various possible substitutions by 1+ elements or co-substitutions by 1+ and 3+ (or >3+) elements further widens the phase space of these materials to a huge number of possible compositions. This opens a new field in the study of functional oxide materials.

In the following we will discuss in particular three compounds with the compositions (Mg,Ni,Co,Cu,Zn)O noted HEOx-0, $(Mg,Ni,Co,Cu,Zn)_{0.95}Li_{0.05}O$ noted HEOx-Li5 and (Li, Mg,Ni,Cu,Zn)O noted thereafter HEOx-Li16.

In order to get a better insight of the charge compensation mechanism in these materials, an XPS study has been performed. For HEOx-0 sample XPS spectrum (survey in Supporting Figure 2) confirmed that all the cations are in a 2+ valence state. All three compounds (HEOx-0, HEOx-Li5 and HEOx-Li16) show identical Mg(1s), Ni(2p), Cu(2p) and Zn (2p) spectra. The effect of $Li^+$ substitution can directly be observed in the spectra, for the doped samples a Li(1s) peak appeared at 54.7 eV whose intensity increased with the amount of lithium (inset in Supporting Figure 3). Notable changes were also observed in the spectra of Co and O. In the undoped sample (HEOx-0) the XPS spectra for the Co 2p peak is typical of a $Co^{2+}$ state, with broad main lines $2p_{3/2}$ and $2p_{1/2}$ and with intense satellite peaks at 786 and 803 eV (Supporting Figure 4).

The $Co^{3+}$ spectra have sharper main lines, located at slightly lower binding energies than $Co^{2+}$ and with a satellite at ~789 eV ("shake-up", O to Co charge transfer) which is particularly used to identify the oxidation state of Co.[14]

The HEOx-Li5 XPS spectrum indicates a decrease of the intensity of $Co^{2+}$ satellite peaks and a shift towards lower binding energies for the main peaks, trend which is clearly visible in the HEOx-Li16 with the apparition of a peak at 789 eV, characteristic of $Co^{3+}$.

It can be concluded from these data that the amount of $Co^{3+}$ increases with the amount of lithium but there is not enough evidence to sustain that *all amount of* cobalt is in 3+ state for the sample (Li,Mg,Co,Ni,Cu,Zn)O. However, it can be confidently affirmed that the mechanism of charge compensation upon $Li^+$ substitution involves, at least in part, the oxidation of cobalt to its 3+ state. This observation can be correlated to the monotonic evolution with the amount of lithium of the intensity of one of the absorption features in the UV-Vis absorption spectra (Supporting Figure 5). These two results further confirm that Li+ unambiguously enters within the rocksalt structure.

A second mechanism for charge compensation might correspond to the formation of oxygen defects. XPS spectra for oxygen, which should be interpreted with caution,[15] show small differences between these three samples (HEOx-0, HEOx-Li5 and HEOx-Li16). The oxygen containing species adsorbed to the surface were removed by low energy Ar+ bombardment and this process is confirmed by the concomitant removal of carbon contamination species. The main O(1s) peak appeared at 529.6 eV assigned to $O^{2-}$ of the lattice (529.7 eV) in addition to a small peak at 531 eV (Supporting Figure 3), whose intensity changes for the three samples and which can be assigned to hydroxides or oxygen defects[16] an "unusual oxygen",[17] lower electron density oxygen[18] described as $O^-$ or with subsurface defects.[16] Thus, a compensation mechanisms involving oxygen cannot be excluded, in agreement with the evolution of the lattice parameters with $Li^+$ substitution, which suggested the existence of two distinct charge compensation mechanisms. It further confirms that this system can be chemically versatile, and that many p-type substitutions and co-substitutions are possible.

As mentioned previously, nothing is known to date about the properties of this new class of materials, and they cannot be easily guessed from those of the constituting elements. Actually, transport properties revealed that HEOx-0 had large electrical resistance at room temperature but presented a very large capacitance with a strong temperature and frequency dependence. For all HEOx samples the resistance changes in an exponential manner as in a semiconductor with electrical band-gaps of about 1eV, gap that decreases with Li substitution (see Supporting Table 2). This band gap can be correlated to the second absorption mechanism that can be observed in the UV-Vis absorption spectra. We note that the large difference in the room temperature resistance between these samples, more than two orders of magnitude, constitutes a signature of the formation of electroactive defects correlated to $Li^+$ substitution. However, it is also consistent with the occurrence of a charge compensation mechanism when substituting the divalent elements by $Li^+$. Indeed, if no charge compensation mechanism occurred, the HEOx-Li5 and HEOx-Li16 samples would probably exhibit a metallic electrical behaviour, as p-type doping would lead to a large concentration of carrier or to the presence of an impurity band close to the valence band maximum. Interestingly, the relative permittivity ($\varepsilon$) is very large for all samples with no maximum observed in the temperature range of the measurement. The relative permittivity at 440 K measured at 20 Hz with a LCR bridge is close to $2 \cdot 10^5$ for HEOx-Li5 (see Supporting Figure 6), making these samples colossal dielectric constant (CDC) materials. Note that the resistance and capacitance values obtained in a LCR setup do not take into account a specific impedance model, the parameters are obtained by considering a simple RC parallel circuit. The permittivity decreases with increasing frequency, this dependence will be discussed in the following part.

Large values of relative permittivity were obtained for all (MgCoNiCuZn)O-based samples with HEOx structure irrespective of substitutions; it seems that colossal dielectric constants is a characteristic of this family. In order to verify this assumption, one of the samples used for the measurement, HEOx-0, was heated at 700°C for 10 hours and slowly cooled. This slow cooling led to solid-solid transformation into a mixture of oxides (Supporting Figure 7), which presented a relatively low resistance (<30 kOhm at room temperature) and no capacitance. It confirms that the colossal dielectric constant is not a common feature of all oxides containing these divalent cations, but is actually a characteristic of the HEOx phase. The origin of this large permittivity is not yet well understood. Indeed, many different mechanisms can lead to CDC behaviours in oxides, which are not always easily determined.[19]

Typically, it may arise in part from extrinsic Maxwell-Wagner-type effects[20] which originates from charge accumulation at interfaces, including grain boundaries or grain surfaces. As mentioned previously, the densities of these samples are moderate (75-80%), which may affect these results. Therefore, the development of different synthesis processes would be of interest to get a better understanding of the actual contributions of the intrinsic and extrinsic mechanisms.

However, as it will be seen in the following, there are "intrinsic" contributions with colossal permittivity that can be determined by high frequency impedance measurements.

Complex impedance measurements showed, on the obtained Bode diagrams (phase in function of the frequency), two distinct dielectric relaxations (Supporting Figure 8). The fitting of all impedance spectra was thus performed by using two R//Q electrical circuits (where Q stands for a pseudo-capacitance) in series with a contact resistance for modelling the contacts and wires resistances, whose value was always < 10 Ohm, in our case (the equivalent circuit and the fitting results are given in Supporting Table 2). It is interesting to note that the first *intrinsic* resistance (*i.e.* not related with the contact resistance) is always lower than the second one (at any temperature), enlightening that the first dielectric relaxation corresponds to a less resistive mechanism (and/or concerns a smaller part of the sample studied). Moreover, the capacitance of this first phenomenon is almost independent of the temperature. On the contrary, the second mechanism has larger resistance (either the process is more resistive or it concerns a larger part of the sample) and exhibits large variation of its capacitance with the temperature. Besides, the lithium content in the sample has a significant impact on the electrical response. Indeed, HEOx-0 exhibits the largest intrinsic (*i.e.* bulk) resistances, whereas the sample with the largest Li content exhibits the lowest intrinsic resistance. Thus, it is worth noting that bulk resistance of these materials can be easily tailored by the lithium amount added during their synthesis. Plots of both intrinsic resistances as a function of the inverse of temperature gave activation energies for these processes (Supporting Table 2). The first relaxation mechanism has a slightly larger value of activation energy than the second one, for all samples. Further work is required to understand the temperature dependence of the dielectric constant in these materials.

The direct action of lithium, in this class of materials, can also be seen through the value of the distortion parameter α, extracted from the fit of the pseudo-capacitance Q of the second contribution (the capacitance of first contribution having an invariable α parameter, close to 1, an ideal plane capacitor). Without Li, this parameter remains also equal to unity in a large range of temperature, whereas it noticeably decreases with the lithium content and becomes sensitive to temperature increase. This is probably related to a larger disorder of the structure, due to the defects that originates from Li substitution, leading to a frequency dispersion of the second relaxation phenomena (deviation from ideal plane capacitor). By correcting the value of the pseudo-capacitance with the α parameter estimated from the fit, we can calculate the second contribution "ideal" permittivity. It is here very interesting to remark that the lithium content has again a large impact on the observed values. The less lithium the sample contains, the higher is its intrinsic resistance (as aforementioned) and the higher is also its maximum permittivity. Moreover, this maximum is observed at lower temperature, when the amount of lithium in the material is increasing, which allows, once again, to tailor the dielectric behaviour of these compounds by adapted/controlled lithium doping, opening a wide variety of different uses for example in energy storage or electronic devices.

The representation of the real part of the global "not ideal" permittivity (deduced from the raw recorded complex impedance data), as well as the evolution of dielectric loss (tan δ) in function of the frequency, gives some information about the dielectric behaviour of the materials in real/practical conditions. We can observe that the permittivity remains unusually high and stable on a wide range of frequencies (2.3MHz-100Hz), especially for low lithium content samples, and noticeably increases at lower frequencies (<10 Hz) to finally reach large values under DC solicitation. Data for HEOx-0 and for HEOx-Li16 are shown in **Figures 2a -d** (more details and data for other compositions are given in Supporting Figure 9 and Supporting Table 2). It is noteworthy that ε remains larger than 1000 in the MHz range while keeping a low dielectric loss, of the order of 0.01. We underline that measurements with different blocking electrodes (Pt/sample/Pt, Ti/sample/Ti, Cu/sample/Cu) gave similar values of relative permittivity.

The more resistive is the compound, the smaller is the *tan δ* value, which is consistent with minimizing the dielectric loss with higher intrinsic resistance.

Concerning the origin of this large relative permittivity we can advance several hypothesis consistent with these materials: dipolar fluctuations in nano-size domains,[21] interface effects or

Maxwell Wagner polarisation,[20] electronic phase separation,[22] defect dipoles,[23] dipoles relaxations[23] or hopping transport, but much further work will be required to clarify these aspects.

**4. Conclusions**

We show here that the new entropy-stabilized oxides can be substituted by aliovalent elements with a charge compensation mechanisms, which strongly widens their phase space to a huge number of possible compositions, and probably properties and applications. As an example of these possibilities, we report in this letter their promising electrical properties namely colossal dielectric constants. We emphasis the impressive amount of substitutions possible as well as the possibility of forming new systems with the same approach. There is a large variety of properties that could be discovered in this new class of materials. We point out that this new topic could represents a *paradigm shift* and is opening a completely new research direction in material science.

**5. Acknowledgements**

The authors acknowledge Dr. Céline Byl for optical diffuse reflectance measurements and Tassatit Nait for her help with samples elaboration. (Supporting Information is available online from Wiley InterScience or from the author)).

Figure 1.

Lattice parameters for a series of compounds in the HEOx family. A linear fit up to 16%Li is shown for (Mg,Co,Ni,Cu,Zn)$_{1-x}$Li$_x$O series, a change in the slope can be observed for concentrations over 16% in lithium. Error bars are smaller than the points.

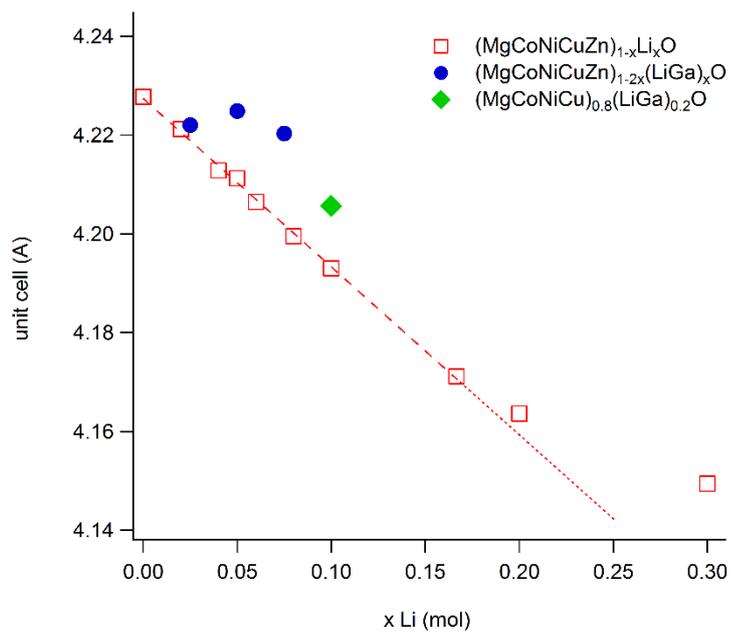

Figure 2.

Real part and imaginary part of the dielectric constant for HEOx-0 (a, b) and HEOx-Li16 (c, d).

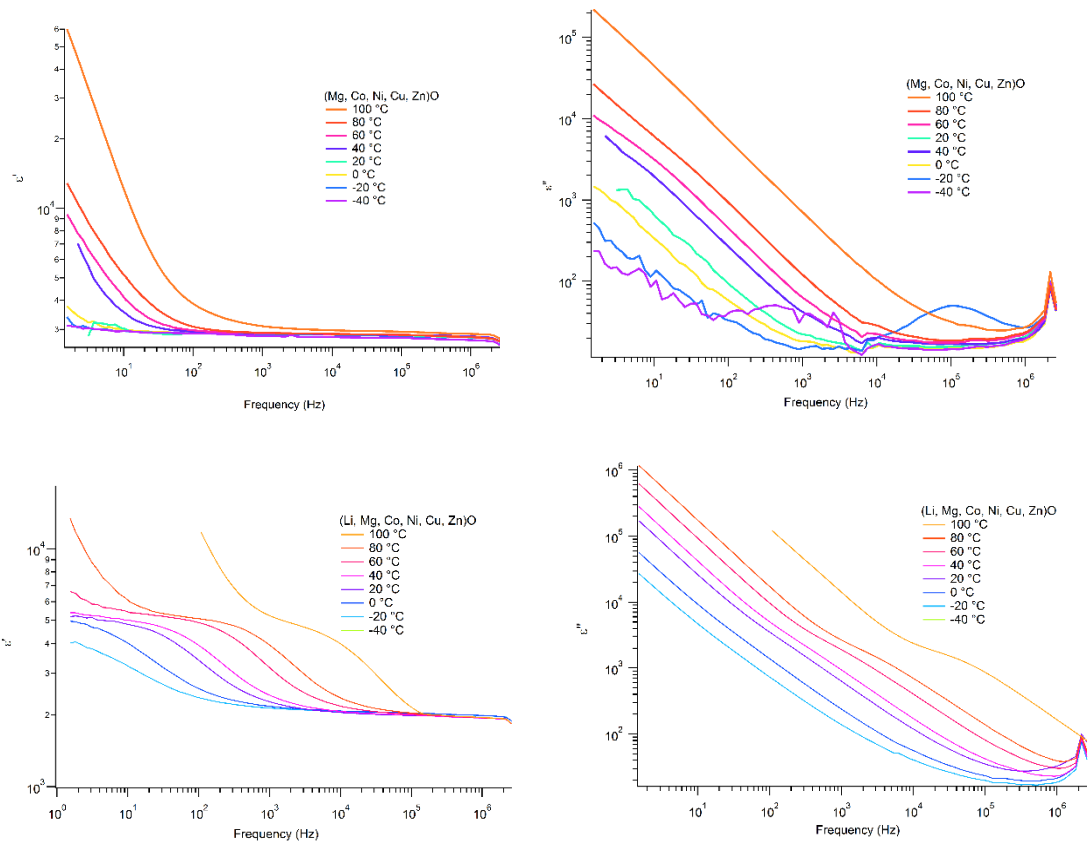

**The table of contents entry**
The synthesis of several new compounds stabilized by high entropy is described. These materials show, for these particular compositions, colossal dielectric constants.

*David Bérardan\*, Sylvain Franger, Diana Dragoe, Arun Kumar Meena and Nita Dragoe\**

Colossal dielectric constant in high entropy oxides

Keywords: (high entropy, colossal dielectric constants)

ToC figure

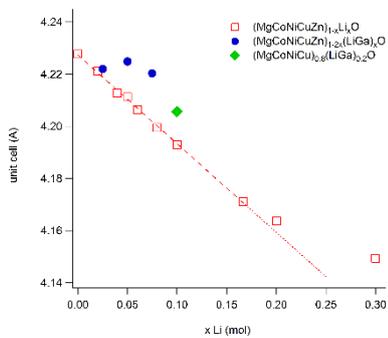

# Supporting Information

**Colossal dielectric constant in high entropy oxides**

By *David Bérardan\**, *Sylvain Franger*, *Diana Dragoe, Arun Kumar Meena and Nita Dragoe\**

[*]   Dr. David Bérardan, Prof. Nita Dragoe
E-mail: david.berardan@u-psud.fr, nita.dragoe@u-psud.fr
ICMMO (UMR 8182 CNRS), Université Paris-Sud, Université Paris-Saclay, 91405 Orsay, 91405 (France)

Prof. Sylvain Franger, Dr. Diana Dragoe, Mr. Arun Kumar Meeta
ICMMO (UMR 8182 CNRS), Université Paris-Sud, Université Paris-Saclay, 91405 Orsay, 91405 (France)

**Supporting Figure 1.**

Powder XRD of selected compositions, for lattice parameters see Supporting Table 1.

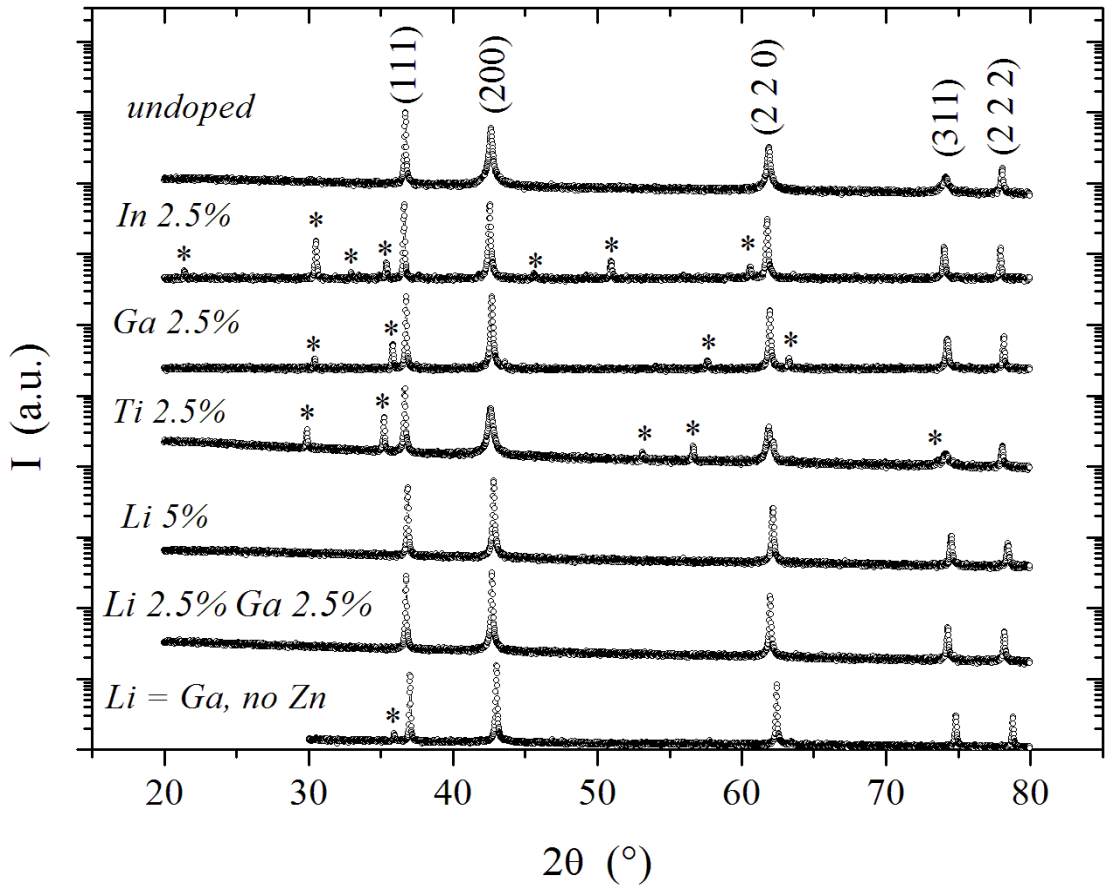

**Supporting Figure 2.**

XPS survey spectra for HEOx-0, HEOx-Li5 and HEOx-Li16.

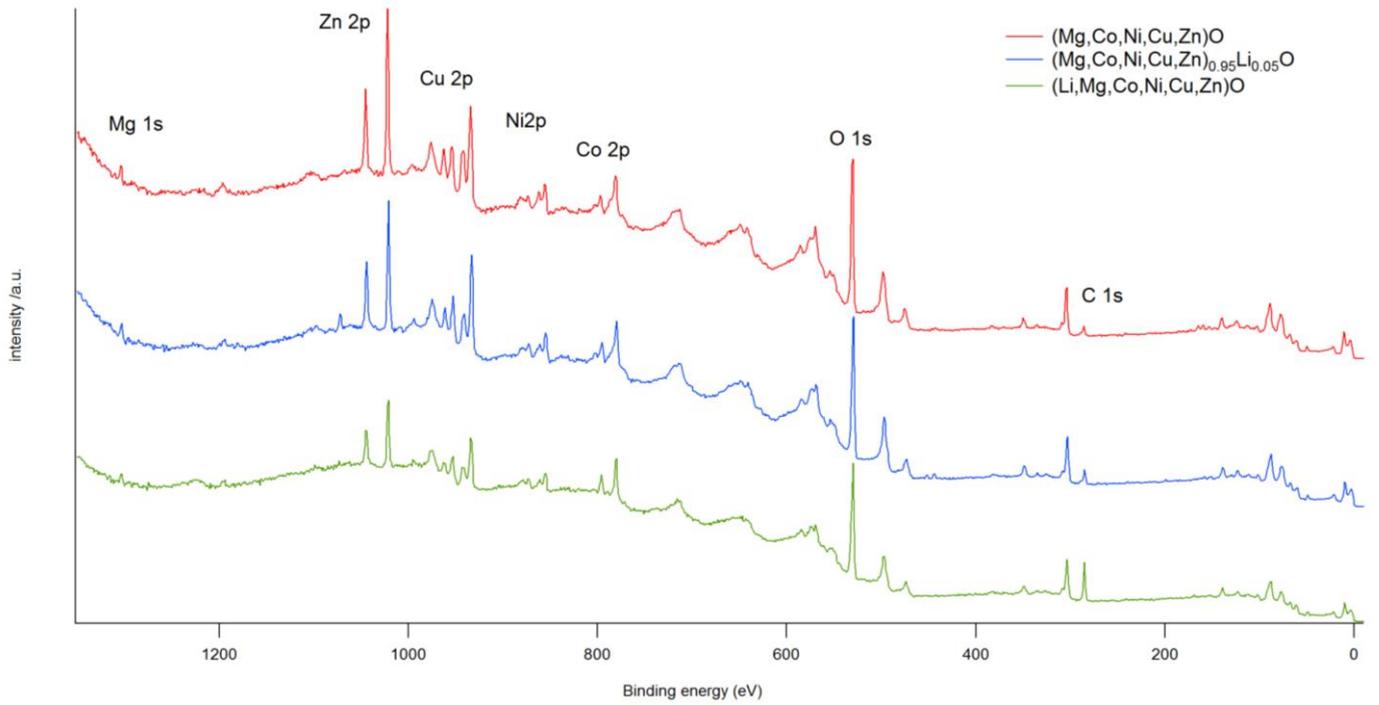

**Supporting Figure 3**

O1s XPS spectra for HEOx A, B and C (inset Li 1s region), the spectra are shifted for clarity.

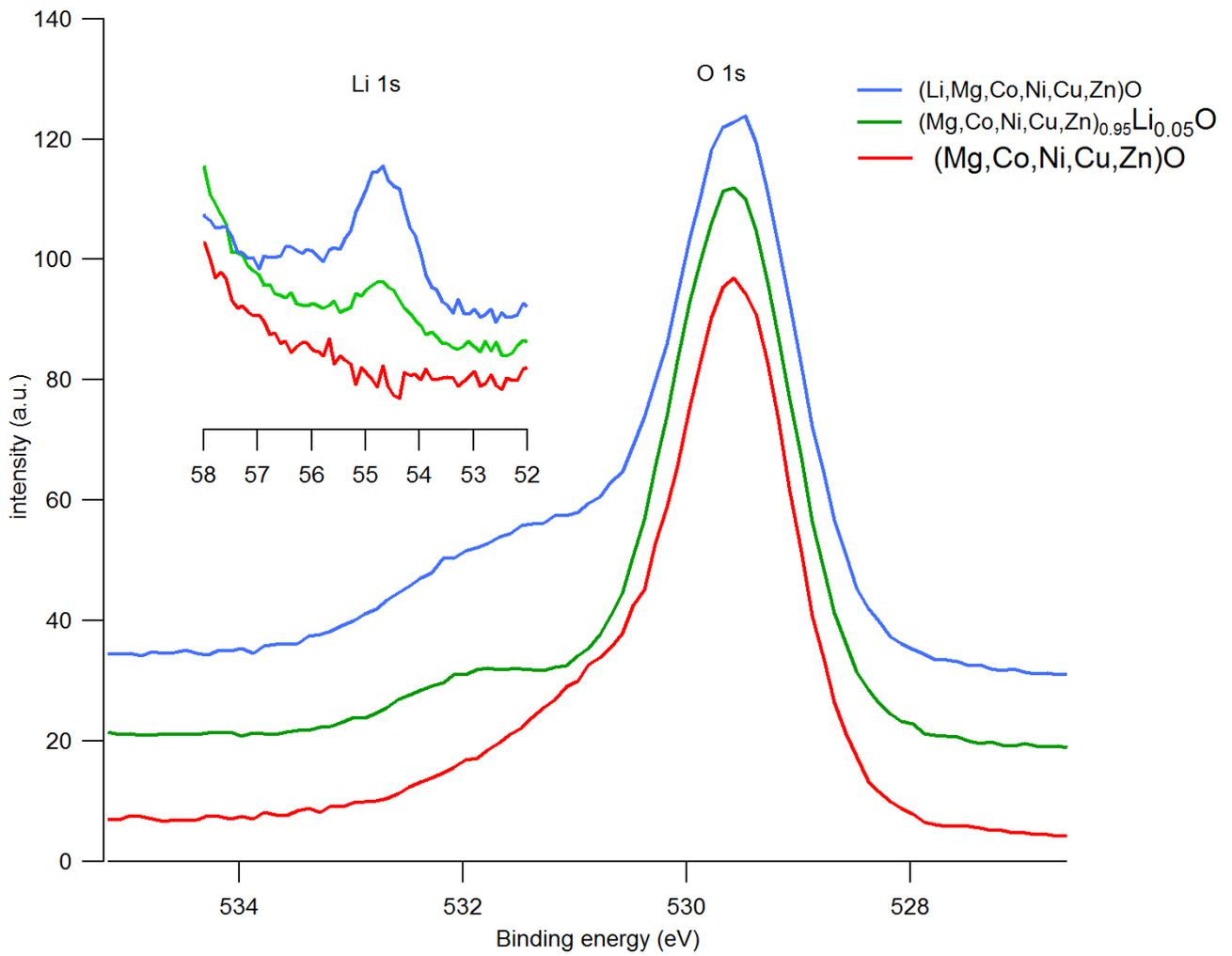

**Supporting Figure 4**

Co-2p core level XPS spectra for HEOx-0 (A), HEOx-Li5 (B) and HEOx-Li16 (C), the main peaks are indicated together with "satellites"; a charge transfer characteristic for $Co^{3+}$ is indicated by an arrow.

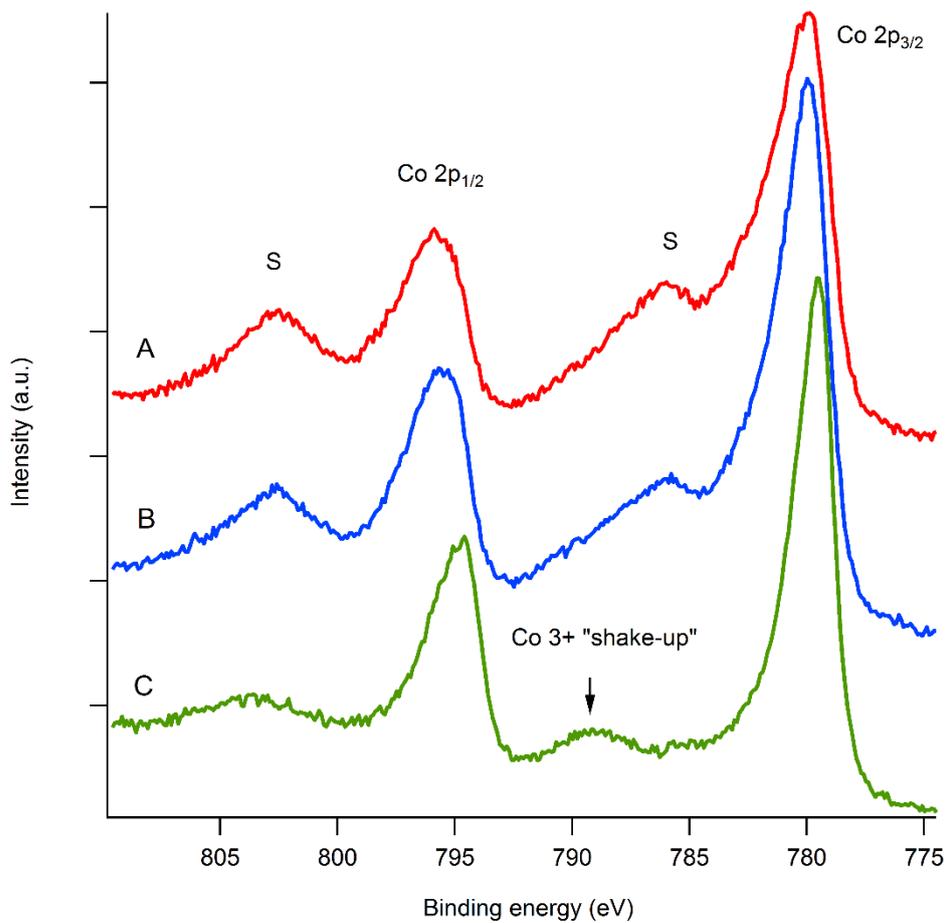

**Supporting Figure 5.**

UV VIS absorption spectra for a series of $(MgCoNiCuZn)_{1-x}Li_xO$.

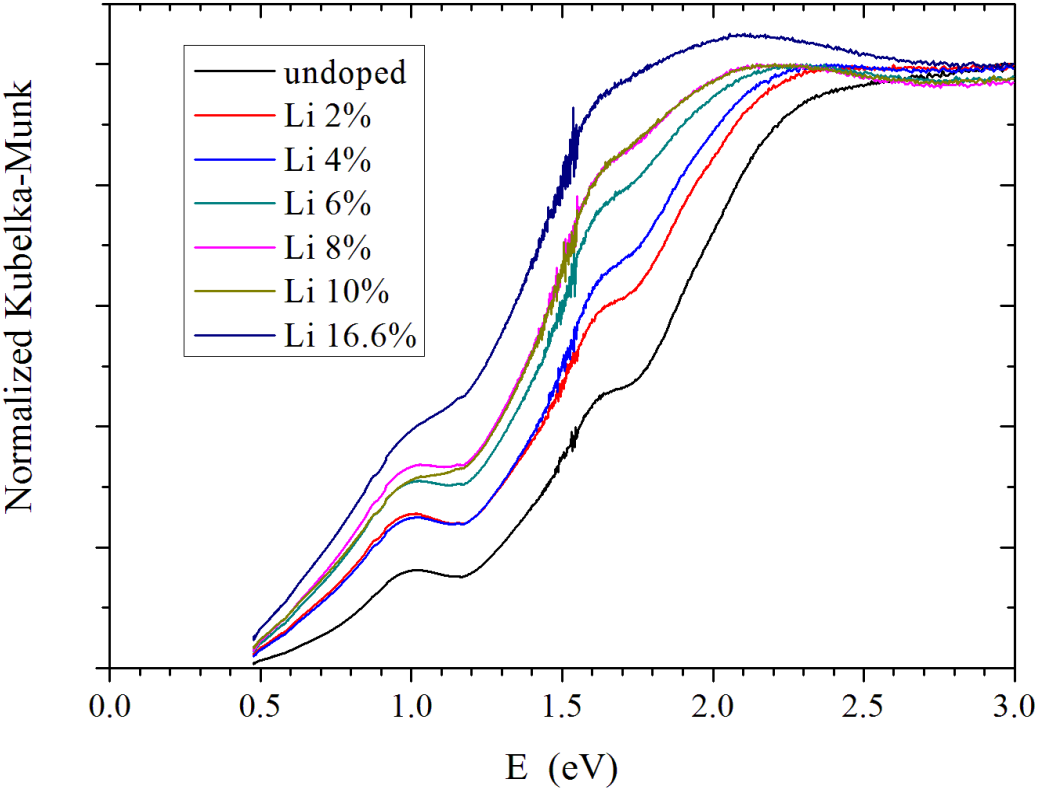

**Supporting Figure 6.**

Resistance and relative permittivity as a function of temperature for HEOx-A-5%Li, measured at low frequencies with a LCR bridge.

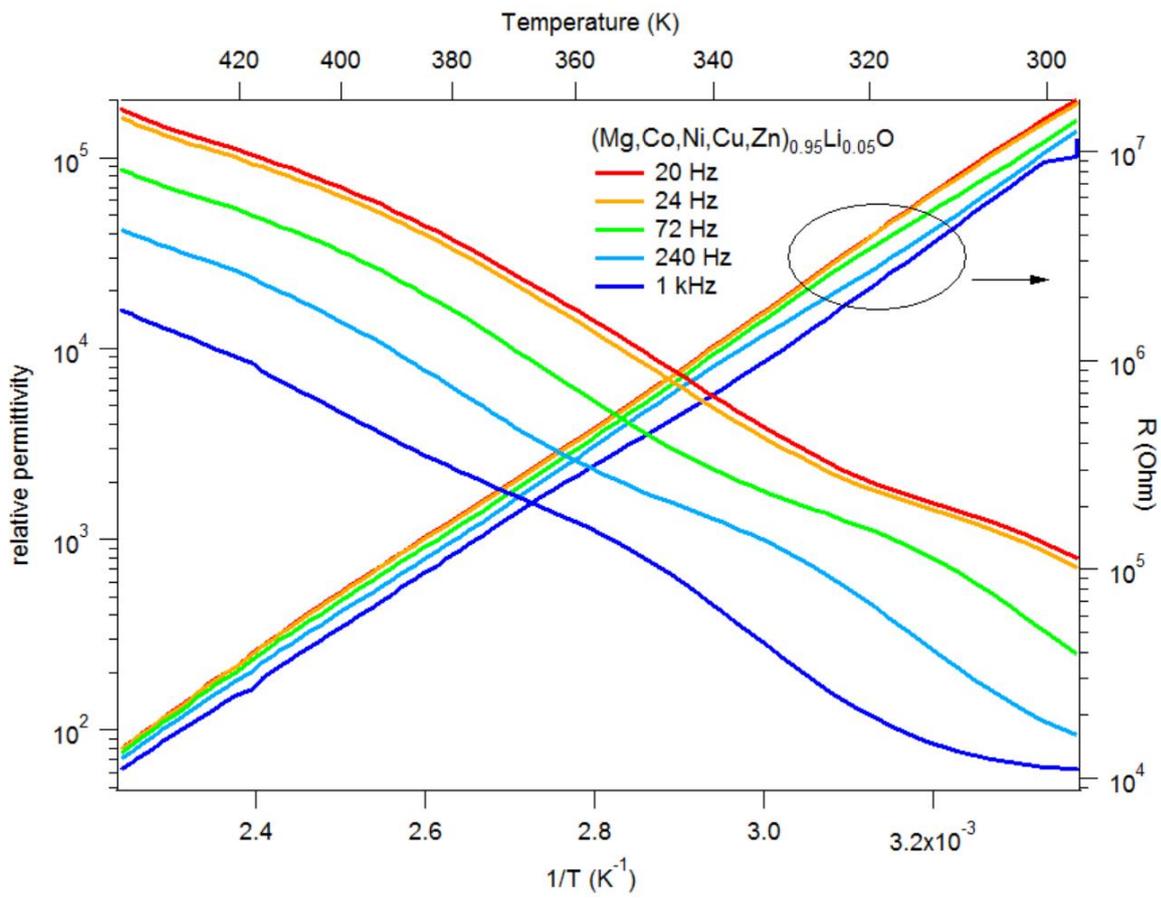

**Supporting Figure 7**

Powder XRD for (MgCoNiCuZn)O after quenching (up) and after slow cooling (bottom).

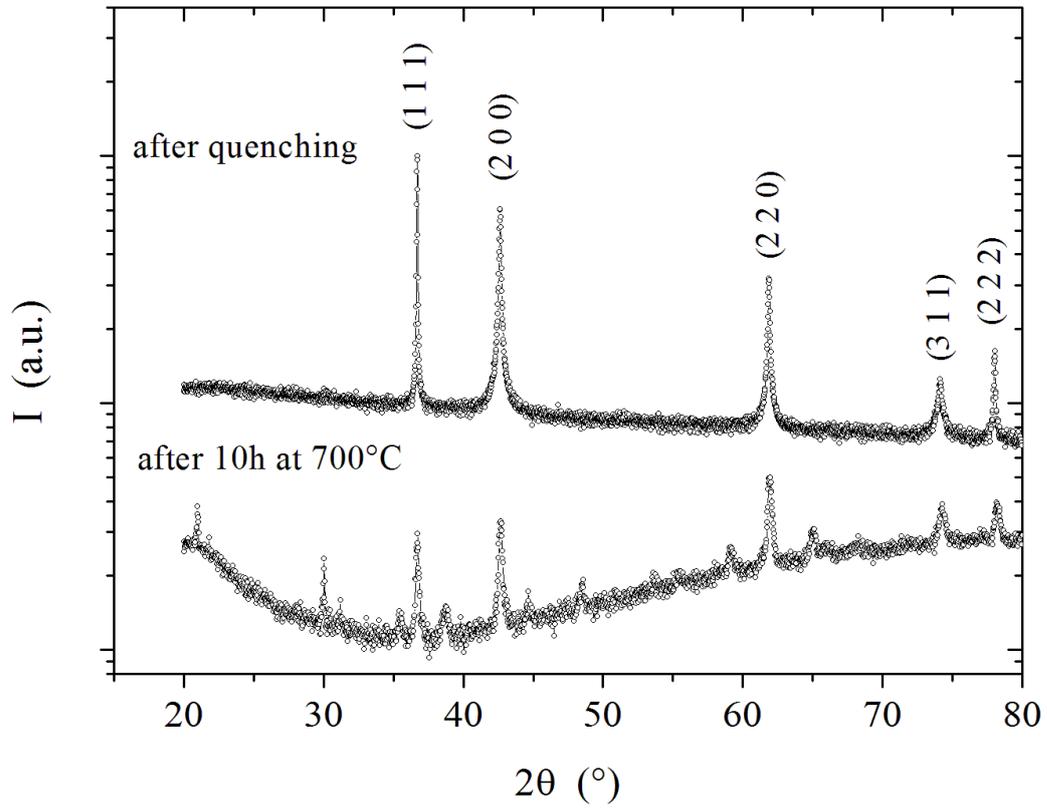

**Supporting Figure 8**

Phase as a function of frequency for HEOx-0.

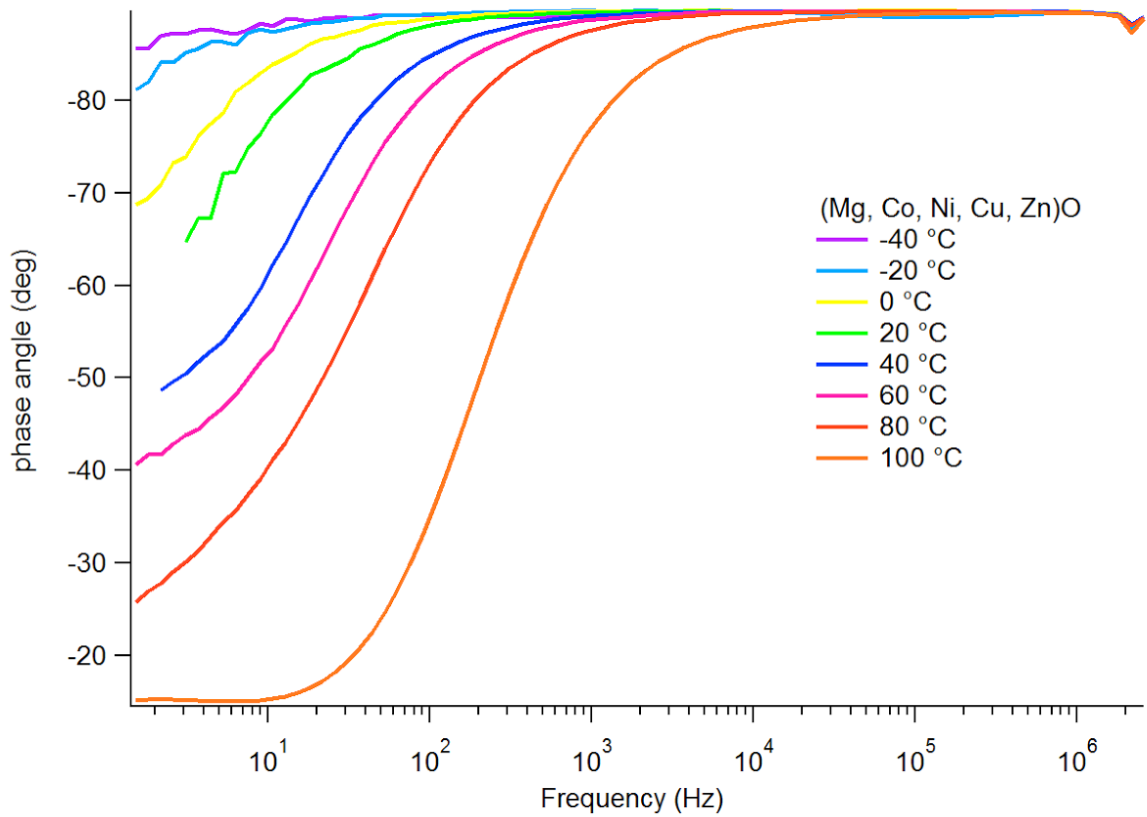

**Supporting Figures 9 a-e**

a. Tan(d) function of frequency at different temperatures for HEOx-0

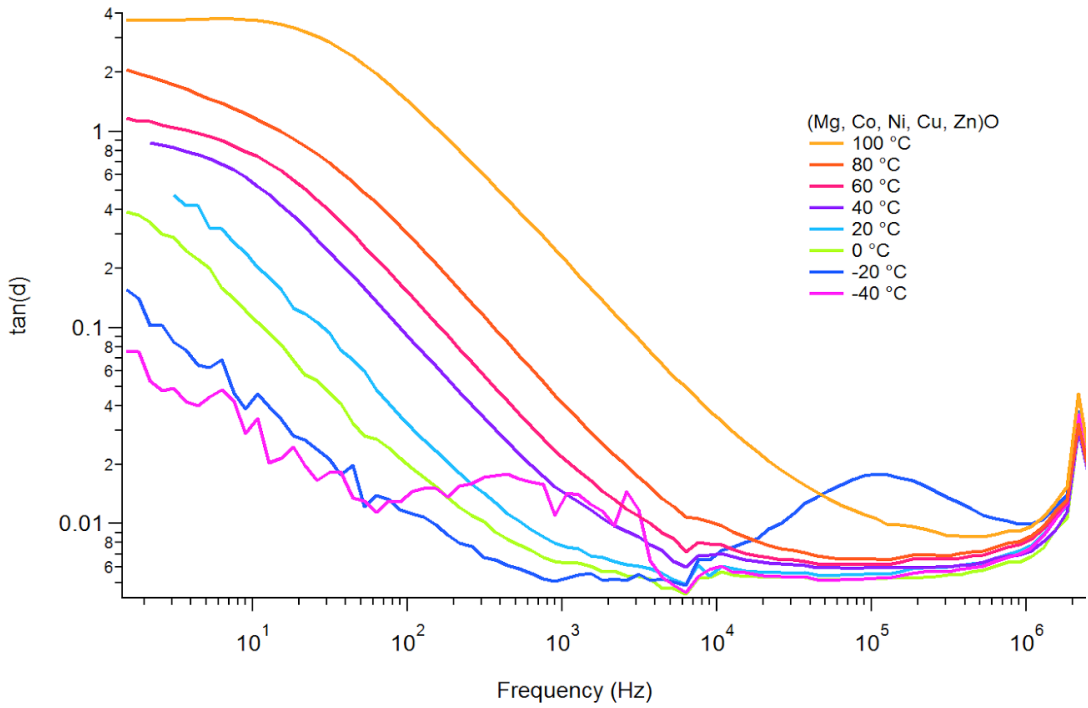

b. Real part of relative permittivity as a function of frequency at different temperatures for HEOx-Li5

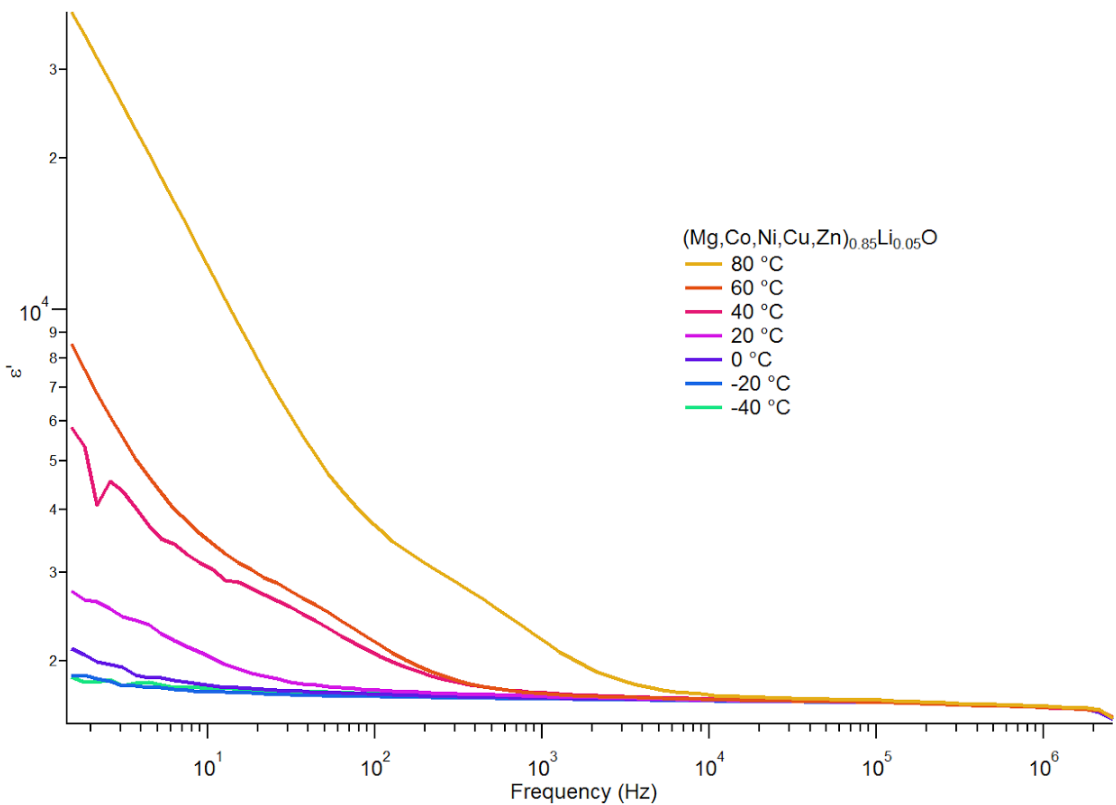

c. Imaginary part of relative permittivity as a function of frequency at different temperatures for HEOx-Li-5

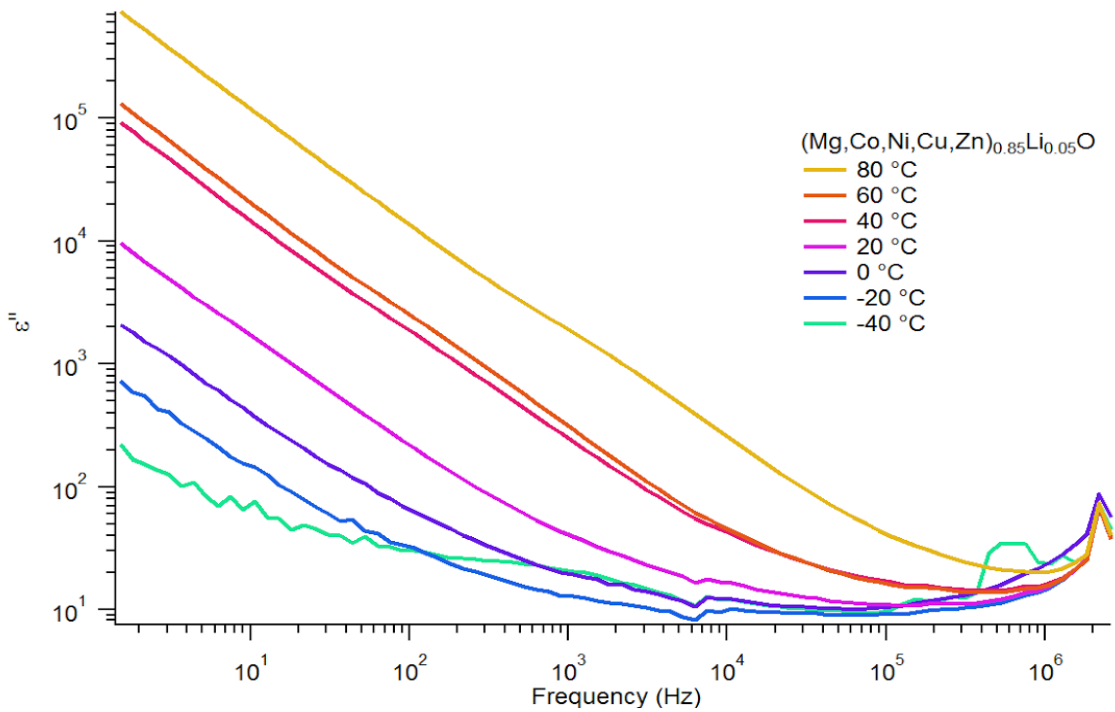

d. Tan(d) as a function of frequency at different temperatures for HEOx-Li5

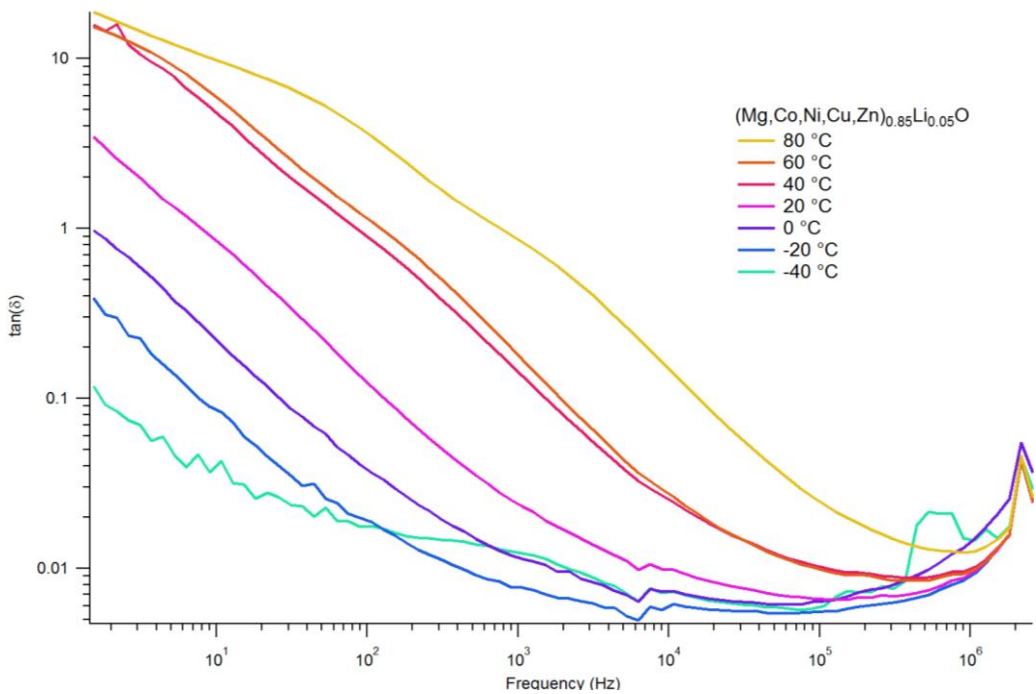

e. Tan (d) as a function of frequency at different temperatures for HEOx-Li16

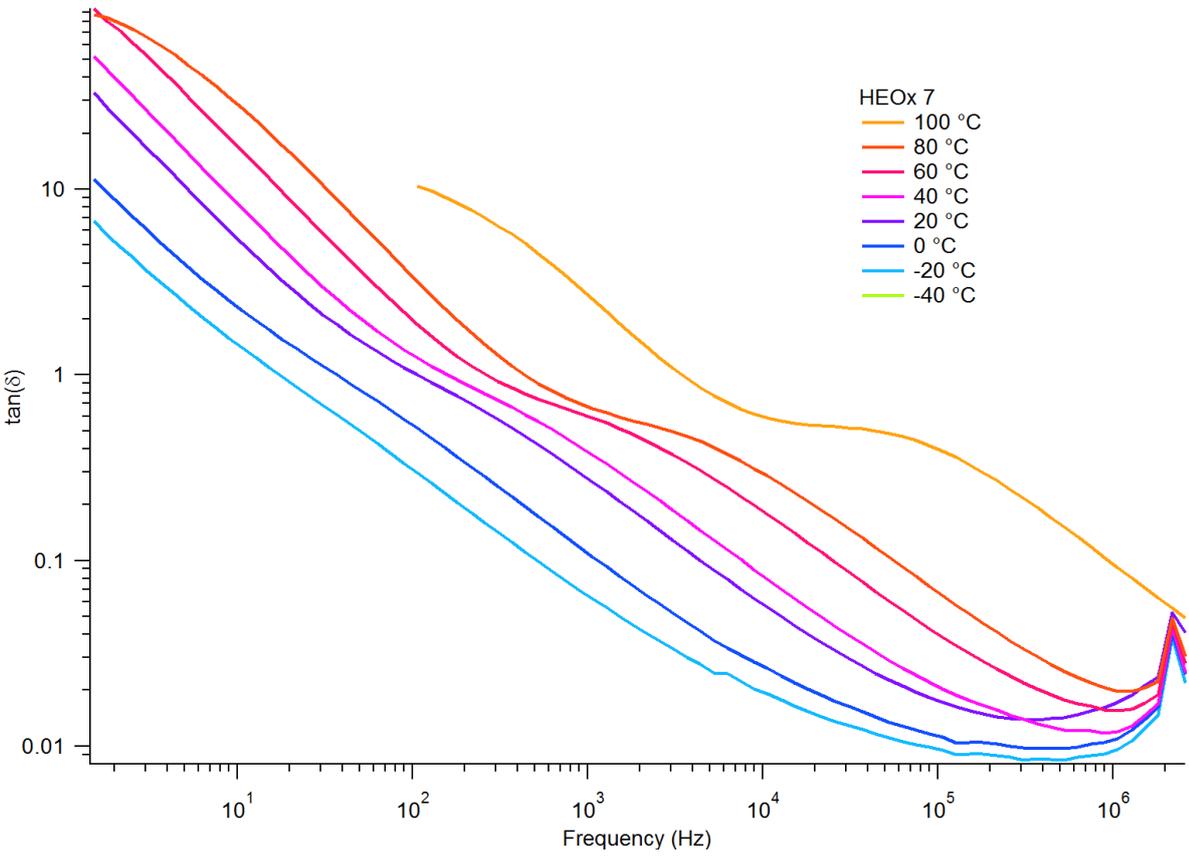

Supporting Table 1

Compositions and lattice parameters of doped Heox

| Nominal composition | result |
|---|---|
| $Mg_{0.2}Co_{0.2}Ni_{0.2}Cu_{0.2}Zn_{0.2}O$ | single phase, a = 4.2277(5)Å |
| $(Mg_{0.2}Co_{0.2}Ni_{0.2}Cu_{0.2}Zn_{0.2})_{0.95}In_{0.05}O$ | mixture of rocksalt, $In_2O_3$ and $In_2MgO_4$ |
| $(Mg_{0.2}Co_{0.2}Ni_{0.2}Cu_{0.2}Zn_{0.2})_{0.95}Ga_{0.05}O$ | mixture of rocksalt and $CuGa_2O_4$ |
| $(Mg_{0.2}Co_{0.2}Ni_{0.2}Cu_{0.2}Zn_{0.2})_{0.95}Li_{0.05}O$ | single phase, a = 4.2113(5)Å |
| $(Mg_{0.2}Co_{0.2}Ni_{0.2}Cu_{0.2}Zn_{0.2})_{0.95}Ti_{0.05}O$ | Mixture of rocksalt and $Co_2TiO_4$ |
| $Mg_{1/6}Co_{1/6}Ni_{1/6}Cu_{1/6}Zn_{1/6}Li_{1/6}O$ | Single phase, a = 4.1712(5) Å |
| $(Mg_{0.2}Co_{0.2}Ni_{0.2}Cu_{0.2}Zn_{0.2})_{0.98}Li_{0.02}O$ | Single phase, a = 4.2213(5) Å |
| $(Mg_{0.2}Co_{0.2}Ni_{0.2}Cu_{0.2}Zn_{0.2})_{0.96}Li_{0.04}O$ | Single phase, a = 4.2129(5) Å |
| $(Mg_{0.2}Co_{0.2}Ni_{0.2}Cu_{0.2}Zn_{0.2})_{0.94}Li_{0.06}O$ | Single phase, a = 4.2065(5) Å |
| $(Mg_{0.2}Co_{0.2}Ni_{0.2}Cu_{0.2}Zn_{0.2})_{0.92}Li_{0.08}O$ | Single phase, a = 4.1995(5) Å |
| $(Mg_{0.2}Co_{0.2}Ni_{0.2}Cu_{0.2}Zn_{0.2})_{0.90}Li_{0.10}O$ | Single phase, a = 4.1930(5) Å |
| $(Mg_{0.2}Co_{0.2}Ni_{0.2}Cu_{0.2}Zn_{0.2})_{0.80}Li_{0.20}O$ | Single phase, a = 4.1636(5) Å |
| $(Mg_{0.2}Co_{0.2}Ni_{0.2}Cu_{0.2}Zn_{0.2})_{0.70}Li_{0.30}O$ | Single phase, peak broadening, a = 4.1494(5) Å |
| $(Mg_{0.2}Co_{0.2}Ni_{0.2}Cu_{0.2}Zn_{0.2})_{0.95}Li_{0.025}Ga_{0.025}O$ | Single phase, a = 4.2221(5) Å |
| $(Mg_{0.2}Co_{0.2}Ni_{0.2}Cu_{0.2}Zn_{0.2})_{0.90}Li_{0.05}Ga_{0.05}O$ | Single phase, a = 4.2249(5) Å |
| $(Mg_{0.2}Co_{0.2}Ni_{0.2}Cu_{0.2}Zn_{0.2})_{0.85}Li_{0.075}Ga_{0.075}O$ | Rocksalt a = 4.2204(5) Å, trace of spinel phase (<2%) |
| $Mg_{0.2}Co_{0.2}Ni_{0.2}Cu_{0.2}Li_{0.1}Ga_{0.1}O$ | Rocksalt a = 4.2057(5) Å, trace of spinel phase (<2%) |

Supporting Table 2

Fitting parameters for the impedance analysis. The equivalent circuit used was:

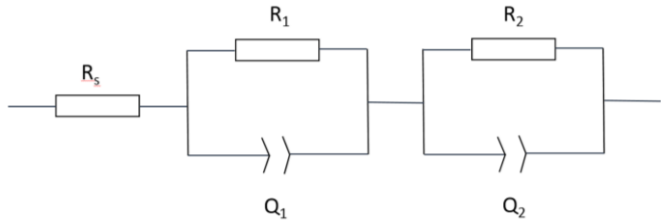

The pseudo-capacitance fit was done using the impedance form of a Constant Phase Element (CPE):

$$Z_{CPE} = \frac{1}{Q(j\omega)^\alpha}$$

| (Mg, Co, Ni, Cu, Zn) O | 40°C | 60°C | 80°C | 100°C |
|---|---|---|---|---|
|  | (fit ± 2%) | (fit ± 3%) | (fit ± 5%) | (fit ± 7%) |
| R1 (MΩ) | 28 | 21 | 12 | 2 |
| Q1 (pF) | 460 | 400 | 370 | 370 |
| R2 (MΩ) | 150 | 86 | 30 | 2 |
| ρ2 (MΩ.cm) | 146 | 84 | 29 | 2 |
| Q2 (pF) | 580 | 750 | 960 | 1320 |
| α | 1 | 1 | 1 | 0.99 |
| C2 (pF) corrected | 580 | 750 | 960 | 1250 |
| ε (S/t = 9.74 mm) | 6850 | 8750 | 11200 | 14550 |

**Ea (contribution 1) = 0.80 eV**
**Ea (contribution 2) = 0.74 eV**

| (Mg, Co, Ni, Cu, Zn)$_{0.95}$Li$_{0.05}$ O | 20°C (fit ± 5%) | 40°C (fit ± 3%) | 60°C (fit ± 3%) | 80°C (fit ± 7%) |
|---|---|---|---|---|
| **R1 (MΩ)** | 30 | 3 | 2 | 0.3 |
| **Q1 (pF)** | 390 | 370 | 350 | 370 |
| **R2 (MΩ)** | 46 | 5 | 3 | 0.6 |
| **ρ2 (MΩ.cm)** | 78 | 8 | 5 | 1 |
| **Q2 (pF)** | 880 | 1030 | 1250 | 1260 |
| **α** | 0.99 | 0.98 | 0.97 | 0.96 |
| **C2 (pF) corrected** | 850 | 930 | 1050 | 930 |
| **ε** (S/t = 16.8 mm) | 5770 | 6275 | 7100 | 6200 |

**Ea (contribution 1) = 0.65 eV**
**Ea (contribution 2) = 0.6 eV**

| (Li, Mg, Co, Ni, Cu, Zn) O | -40°C (fit ± 5%) | -20°C (fit ± 3%) | 0°C (fit ± 2%) | 20°C (fit ± 3%) |
|---|---|---|---|---|
| R1 (MΩ) | 16 | 7 | 4 | 1 |
| Q1 (pF) | 530 | 530 | 470 | 480 |
| R2 (MΩ) | 52 | 24 | 11 | 4 |
| ρ2 (MΩ.cm) | 73 | 34 | 15 | 5 |
| Q2 (pF) | 610 | 670 | 880 | 910 |
| α | 0.99 | 0.98(7) | 0.98(1) | 0.97 |
| C2 (pF) corrected | 590 | 630 | 800 | 750 |
| ε (S/t = 13.98 mm) | 4800 | 5150 | 6500 | 6200 |

| (Li, Mg, Co, Ni, Cu, Zn) O | 40°C (fit ± 4%) | 60°C (fit ± 5%) | 80°C (fit ± 5%) | 100°C (fit ± 6%) |
|---|---|---|---|---|
| R1 (MΩ) | 0.60 | 0.15 | 0.06 | 0.004 |
| Q1 (pF) | 490 | 530 | 550 | 520 |
| R2 (MΩ) | 2.5 | 1.3 | 0.70 | 0.09 |
| ρ2 (MΩ.cm) | 3.5 | 2 | 1 | 0.1 |
| Q2 (pF) | 960 | 970 | 980 | 1400 |
| α | 0.96 | 0.95(5) | 0.95(0) | 0.92 |
| C2 (pF) corrected | 740 | 710 | 670 | 630 |
| ε (S/t = 13.98 mm) | 6050 | 5750 | 5450 | 5150 |

**Ea (contribution 1) = 0,4 eV**
**Ea (contribution 2) = 0,3 eV**